\documentclass[twocolumn, secnumarabic, amssymb, nobibnotes, prapplied, superscriptaddress]{revtex4-2}

\newcommand{\murm}{%
  \ifmmode
    \mathchoice
        {\hbox{\normalsize\textmu}}
        {\hbox{\normalsize\textmu}}
        {\hbox{\scriptsize\textmu}}
        {\hbox{\tiny\textmu}}%
  \else
   ~\textmu
  \fi
}

\usepackage{bbold}

\usepackage{lipsum}
\usepackage{amsfonts,amssymb,amsmath}
\usepackage{graphicx}
\usepackage{bm}
\usepackage{calc}
\usepackage[x11names]{xcolor}
\usepackage{siunitx}
\usepackage{tikz}
\usepackage{wasysym}
\usepackage{dsfont}
\usepackage{hyperref}

\newcommand{\fmarki}{*}
\newcommand{\fmarkii}{\ensuremath{\dagger}}
\newcommand{\fmarkiii}{\ensuremath{\ddagger}}

\def\@fnsymbol#1{{\ifcase#1\or \fmarki\or \fmarkii\or \fmarkiii\or \fmarkiv\or \fmarkv\or \fmarkvi\or \fmarkvii\or \fmarkviii\or \fmarkix \else\@ctrerr\fi}}
\makeatother

\renewcommand{\fmarki}{b$_1$}
\renewcommand{\fmarkii}{b$_2$}
\renewcommand{\fmarkiii}{c$_3$}

\definecolor{color_comment}{rgb}{0.8, 0.3, 0.3}

\definecolor{color_out}{rgb}{0.7, 0.7, 0.7}

\definecolor{color_new}{rgb}{0.3, 0.8, 0.3}

\newcommand{\nv}{N-$V$}

\definecolor{color_comment}{rgb}{0.8, 0.3, 0.3}

\begin{document}

\title{Gate-set evaluation metrics for closed-loop optimal control on nitrogen-vacancy center ensembles in diamond}

\author{Philipp J. Vetter}
\thanks{These authors contributed equally to this work. \\Corresponding author: philipp.vetter(at)uni-ulm.de}
\affiliation{Institute for Quantum Optics, Ulm University, Albert-Einstein-Allee 11, 89081 Ulm, Germany}
\affiliation{Center for Integrated Quantum Science and Technology (IQST), 89081 Ulm, Germany}
\author{Thomas Reisser}
\thanks{These authors contributed equally to this work. \\Corresponding author: philipp.vetter(at)uni-ulm.de}
\affiliation{Peter Grünberg Institute -- Quantum Control (PGI-8), Forschungszentrum J\"ulich GmbH, D-52425 Germany}
\affiliation{Institute for Theoretical Physics, University of Cologne, D-50937 Germany}
\author{Maximilian G. Hirsch}
\affiliation{Institute for Quantum Optics, Ulm University, Albert-Einstein-Allee 11, 89081 Ulm, Germany}
\affiliation{Center for Integrated Quantum Science and Technology (IQST), 89081 Ulm, Germany}
\affiliation{Current address: NVision Imaging Technologies GmbH, Wolfgang-Paul-Straße 2, 89081 Ulm, Germany}
\author{Tommaso Calarco}
\affiliation{Peter Grünberg Institute -- Quantum Control (PGI-8), Forschungszentrum J\"ulich GmbH, D-52425 Germany}
\affiliation{Institute for Theoretical Physics, University of Cologne, D-50937 Germany}
\affiliation{Dipartimento di Fisica e Astronomia, Università di Bologna, 40127 Bologna, Italy}
\author{Felix Motzoi}
\affiliation{Peter Grünberg Institute -- Quantum Control (PGI-8), Forschungszentrum J\"ulich GmbH, D-52425 Germany}
\affiliation{Institute for Theoretical Physics, University of Cologne, D-50937 Germany}
\author{Fedor Jelezko}
\affiliation{Institute for Quantum Optics, Ulm University, Albert-Einstein-Allee 11, 89081 Ulm, Germany}
\affiliation{Center for Integrated Quantum Science and Technology (IQST), 89081 Ulm, Germany}
\author{Matthias M. Müller}
\affiliation{Peter Grünberg Institute -- Quantum Control (PGI-8), Forschungszentrum J\"ulich GmbH, D-52425 Germany}
%

\date{March 2024}


\begin{abstract}
A recurring challenge in quantum science and technology is the precise control of their underlying dynamics that lead to the desired quantum operations, often described by a set of quantum gates. 
These gates can be subject to application-specific errors, leading to a dependence of their controls on the chosen circuit, the quality measure and the gate-set itself. 
A natural solution would be to apply quantum optimal control in an application-oriented fashion.
In turn, this requires the definition of a meaningful measure of the contextual gate-set performance. 
Therefore, we explore and compare the applicability of quantum process tomography, linear inversion gate-set tomography, randomized linear gate-set tomography, and randomized benchmarking as measures for closed-loop quantum optimal control experiments, using a macroscopic ensemble of nitrogen-vacancy centers in diamond as a test-bed. 
Our work demonstrates the relative trade-offs between those measures and how to significantly enhance the gate-set performance, leading to an improvement across all investigated methods.

\end{abstract}

\maketitle



\section{Introduction}
\label{sec:introduction}
%
%
Be it quantum computing, information, communication or metrology, precise control over a quantum system is a prerequisite for any successful application.
To ensure robust results and the ability to run long quantum circuits, a great effort is spent by, e.g., IBM and Google on a periodic re-calibration of their quantum chips~\cite{ibm_calibration, michielsen2017benchmarking, papivc2023error, tornow2022minimum, demirdjian2022variational, arute2019quantum}.
To this end a plethora of error correction and error suppression schemes are developed to achieve fault tolerant quantum computing~\cite{kelly2015state, terhal2015quantum, chiaverini2004realization, cory1998experimental, cai2023quantum, endo2018practical}. 
Another example is quantum sensing, where the measured signals can be biased by quantum error correction~\cite{rojkov2022bias} or distorted by faulty gates that are not accounted for~\cite{vetter2022zero}. 

In general, the precision of the system often suffers under, e.g., erroneous operations, wrongly populated states or limited knowledge of the system. 
Moreover, these errors must not necessarily accumulate linearly~\cite{willsch2017gate, lagemann2023fragility, marxer2023implementing}, leading to a dependence of the optimal controls for specific operations on the chosen circuit and the corresponding point of time within the selected circuit~\cite{michielsen2017benchmarking, papivc2023error}.
Thus, the quality of an operation has to be viewed within the context of its application, rather than looking at isolated individual gates; the focus should be shifted to their performance with respect to the chosen circuit and the entire gate-set. 
The ideal scenario would be to find a set of gates that performs universally well, regardless of the planned experiment.
Since such a gate-set will heavily depend on the chosen experimental system, Quantum Optimal Control (QOC)~\cite{dAlessandro2021introduction, brif2010control, glaser2015training, rembold2020introduction, mueller2022chopped} is well suited to tackle this complex problem.

Several optimization algorithms~\cite{khaneja2005optimal, machnes2011comparing, motzoi2011optimal,caneva2011chopped, rach2015dressing,  fortunato2002design,  ciaramella2015newton, leung2017speedup, preti2022continuous} are available in dedicated software packages such as our Quantum Optimal Control Suite (QuOCS)~\cite{rossignolo2023quocs} that allows for closed-loop black-box optimization employing the dCRAB algorithm~\cite{rach2015dressing, mueller2022chopped} via an interface to the lab software Qudi~\cite{binder2017qudi}. 
Such closed-loop optimizations based on measurements on the quantum system have been employed for gate optimization in superconducting qubits via randomized benchmarking~\cite{chow2010optimized, kelly2014optimal}, to optimize the preparation and phase transitions of Bose-Einstein condensates~\cite{heck2018remote, rosi2013fast}, enhance the macroscopic hyperpolarization in pentacene-doped naphthalene crystals~\cite{marshall2022macroscopic} and for autonomous calibration of single-qubit gates as well as robust magnetometry of nitrogen-vacancy (\nv) centers in diamond~\cite{frank2017autonomous, oshnik2022robust}. 
For \nv centers in general, QOC has become a valuable tool for a wide range of challenges~\cite{said2009robust, scheuer2014fast, dolde2014high, poggiali2018optimal, haeberle2013high, noebauer2015smooth, ziem2019quantitative, konzelmann2018robust, mueller2018noise, vetter2022zero, rembold2020introduction}.
Room-temperature accessibility and control of their electronic spin state makes \nv centers ideal sensors for magnetic and electric fields~\cite{balasubramanian2008nanoscale, mamin2013nanoscale, schmitt2017submillihertz, dolde2011electric}.

Ensembles of \nv centers have particularly become a focus of attention, as these allow to drastically improve the sensitivity~\cite{glenn2018high, barry2020sensitivity, balasubramanian2019dc}, create spatially resolved images of, e.g.,~local magnetic fields~\cite{horsley2018microwave, mizuno2018wide, scholten2021widefield} and can even be used to create new phases of matter~\cite{choi2017observation, choi2019probing}. 
Due to their macroscopic size, many individual \nv centers contribute to the measurement, which ensures a very fast signal acquisition but can also result in strong detuning and large amplitude errors.
These errors can negatively influence gate performances especially at different time-scales, as can be seen in Fig.~\ref{fig:Figure_1}~a).
A macroscopic ensemble of \nv centers thus combines the effects of circuit- and length-dependent errors, large state preparation and measurement (SPAM) errors, a non-Markovian noise environment through coupling to other spins and a high sensitivity to possible experimental drifts and distortions of the control pulses, making it an ideal test-bed to investigate the applicability of QOC for these kinds of inhomogeneous error mechanisms.

In order to find a universally well performing gate-set, a good measure is required, that reflects the quality of operations on the system in a holistic framework, being sensitive to the system dynamics on different time-scales and reliably accessible via measurements.
To this end, we derive several measures of our gate-sets performance based on classical Quantum Process Tomography (QPT)~\cite{poyatos1997complete, chuang1997prescription, leibfried1996experimental, frank2017autonomous}, Linear-inversion Gate-Set Tomography (LGST)~\cite{greenbaum2015introduction, blume2013robust} and Randomized Linear Gate-Set Tomography (RLGST)~\cite{gu2021randomized} as well as Optimized Randomized Benchmarking for Immediate Tune-up (ORBIT)~\cite{kelly2014optimal}.
We experimentally asses their applicability for closed-loop optimizations in our test-bed and investigate how they reflect the system dynamics.
We proceed by performing several optimization runs per method and cross-evaluate the performance of the optimized gate-set against all other methods as well as by Randomized Benchmarking (RB)~\cite{knill2008randomized, magesan2011scalable, magesan2012characterizing}. 
Large improvements of the gate-set are observed in almost all optimization scenarios, significantly outperforming the provided guess and even outperforming the commonly used fastest possible rectangular shaped pulse.
Moreover, we investigate how the chosen circuit length influences our obtained results for selected methods.
Finally, we discuss the relative trade-offs of the individual methods applied on these types of systems and show how to best enhance the performance of the selected gate-set to find a universally well performing set.


\section{Results}
\label{sec:results}

The experiments are performed with a CVD-grown diamond with a natural abundance ($1.1~\%$) of ${}^{13}\text{C}$ nuclei and a 10~\textmu m thick layer of \nv centers. 
The layer contains an \nv concentration of roughly 1.5~ppm and is excited by a 532~nm laser with a beam waist of $\approx 38$~\textmu m.
We apply a static magnetic field of 572~G to polarize the inherent nitrogen spin~\cite{jacques2009dynamic} and to create an effective qubit between the $\vert 0\rangle$ and $\vert -1\rangle$ states. 
Gates are generated by microwave pulses sent through a straight microwave antenna made out of gold and placed on top of the diamond. 
%

\subsection{The Figures of Merit}
\label{subsec:fom_intro_and_validation}

\begin{figure*}
    \centering
    \includegraphics[width=0.99\textwidth]{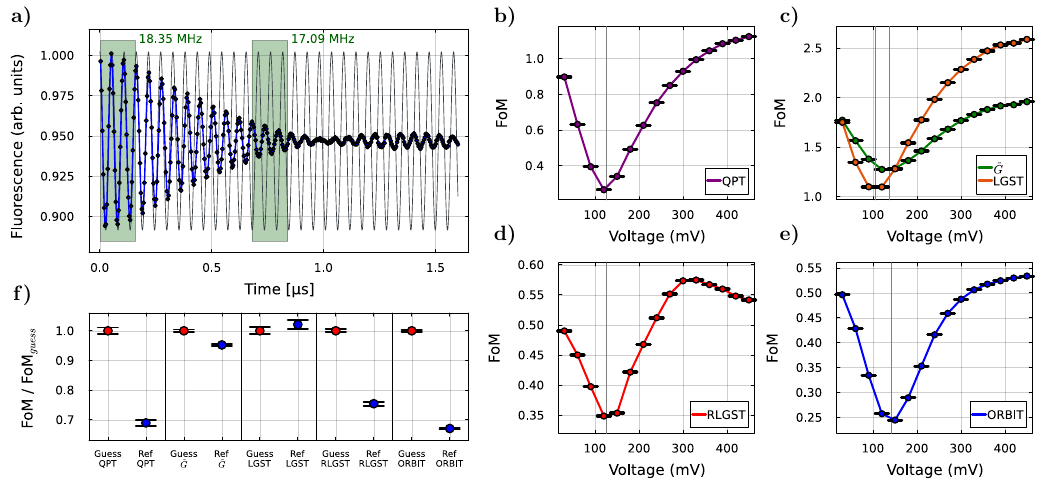}
    \caption{Ensemble dynamics and FoM validity measurements.
    \textbf{a)} Rabi experiment with the \nv center ensemble, showing a strong beating.
    The green areas highlight the time dependence of the Rabi frequency and the black background oscillation visualizes the corresponding shift.
    \textbf{b)} Linear sweep of the guess pulse amplitude for QPT, \textbf{c)} LGST and $\Tilde{G}$, \textbf{d)} RLGST and \textbf{e)} ORBIT.
    The ordinate shows the absolute value of the mean FoM from $20$ measurements and the uncertainty. 
    The minimum of the corresponding FoM is indicated by a grey line.
    \textbf{f)} Comparison of the FoM definitions for the guess (red) and reference pulse (blue), normalized by $\text{FoM}_{\text{guess}}$. 
    Each data point shows the mean value of $20$ measurements with the corresponding standard deviation.}
    \label{fig:Figure_1}
\end{figure*}

We choose the gate-set
\begin{equation}
\begin{split}
        \mathcal{G}& =\lbrace G_0, G_1, G_2, G_3, G_4, G_5, G_6 \rbrace\\
        & = \lbrace \mathds{1}, \mathcal{X}_{\pi/2}, -\mathcal{X}_{\pi/2}, \mathcal{Y}_{\pi/2}, -\mathcal{Y}_{\pi/2}, \mathcal{X}_{\pi}, \mathcal{Y}_{\pi}  \rbrace \,   
    \end{split}
    \label{eq:gate_set}
\end{equation}
for our experiments. 
Here, $\mathds{1}$ is the identity and $\mathcal{X}$ and $\mathcal{Y}$ are the rotations around the $x$- and $y$- axis, where the rotation angle is given by the index.
The gate-set is chosen such that we can generate the full Clifford group~\cite{epstein2014investigating} needed for RB and ORBIT.
We limit our optimizations to $G_2=-\mathcal{X}_{\pi/2}$ to analyze how the change of an individual pulse affects the performance of our gate-set Eq.~\eqref{eq:gate_set} and to ensure that the optimization converges quickly. 
Except for QPT, all methods evaluate the performance of the entire gate-set, leading to the exact same measurements as if we would optimize all gates at once. 
Thus we do not expect a loss of generality of our results by limiting ourselves to one optimized gate. 
The initial state is set to $\vert\rho\rangle\rangle=\vert0\rangle\rangle$ and the POVM to $\vert E\rangle\rangle=\vert0\rangle\rangle$ for all of our experiments.
Note that operators and states are given in Hilbert-Schmidt space to benefit from a simplified syntax (see supplementary information~\cite{supplementary}).

The performance of the gate-set is evaluated through measures based on QPT~\cite{poyatos1997complete, chuang1997prescription, leibfried1996experimental, frank2017autonomous}, LGST~\cite{greenbaum2015introduction, blume2013robust}, RLGST~\cite{gu2021randomized} and ORBIT~\cite{kelly2014optimal}.
Additionally, we consider an adaption of LGST where we only take the targeted expectation values into account and label it $\Tilde{G}$.
The information extracted from these schemes is condensed down to a single real-valued number, called the Figure of Merit (FoM).
The derivation and exact formula of the FoMs for each scheme can be found in the Methods section~\ref{subsec:fom_definitions}. 
Note that the FoM is minimized in all experiments.

During the closed-loop QOC experiments, QuOCS varies the $S_x$ and $S_y$ components of the microwave pulse amplitudes. 
To check the feasibility of our FoM definition, we sweep the pulse amplitude by varying the applied microwave voltage at a constant length $T=30\,$ns for a rectangular pulse, effectively creating a gate $G_2$ that either under- or over-rotates the spin instead of performing a $-\mathcal{X}_{\pi/2}$ operation. 
We select a circuit length of $L_R=18$ for RLGST, which corresponds to a circuit length of $L_O=10$ Cliffords for ORBIT and average over 300 circuits. 
%
%
The effect of this variation on the FoMs is shown in Fig.~\ref{fig:Figure_1}~b)-e). 
All curves show a distinct minimum from which the FoM increases upon deviation from the ideal voltage within the context of the individual technique. 
The obtained results again highlight the problem of circuit- and length-dependence of the optimal control parameters of our system, as we observe different minima, i.e. a different best amplitude, for all methods. 

A second important property for a FoM is that it must properly reflect the dynamics of the quantum system. 
To check this, we compare two rectangular pulses of different length. 
A short pulse with large amplitude is expected to outperform a long pulse with small amplitude because of its increased robustness against detuning errors as well as a smaller decoherence due to the shorter duration.
In Fig.~\ref{fig:Figure_1}~f) the FoM of a 30~ns long rectangular pulse, which we from now on label as the "guess" pulse since it will be later handed to QuOCS as the guess at the start of our closed-loop QOC experiments, and the FoM of the shortest possible rectangular pulse of $\approx 14$~ns, labeled as "reference" pulse, are shown. 
In both cases, the amplitude is determined through a Rabi experiment to match the selected pulse length. 
The shortest pulse is constrained by the maximal voltage that can be generated.  
Except for LGST, all methods show the expected behaviour, that the short reference pulse achieves a significantly lower FoM than the guess pulse. 
While $\Tilde{G}$ shows a smaller difference when compared with QPT, RLGST and ORBIT, the observed difference is well above the margin of error. 
LGST indicates that the reference pulse performs slightly worse than the guess pulse and it appears that the method cannot be used for our macroscopic ensemble as it does not capture the expected dynamics of our effective spin correctly. 
Therefore, we drop LGST as a measure for the cross-comparison of the optimized gate-sets between the different methods but still explore how it competes as a FoM for closed-loop optimizations.
%

\subsection{Optimization Gain}
\label{subsec:gain}

When comparing different analysis methods with each other, it is important to not only consider the absolute change in their FoM but to compare the relative change between two fixed reference points in order to correctly interpret the measurement results.
Therefore, we introduce the so-called gain
\begin{equation}
    \text{Gain} = \frac{\text{FoM}_\text{opt}-\text{FoM}_\text{guess}}{\text{FoM}_\text{ref}-\text{FoM}_\text{guess}}
    \label{eq:optimization_gain}
\end{equation}
which displays the achieved improvement of the $\text{FoM}_\text{opt}$ in relation to the FoM of the reference and guess pulse. 
If the gain exceeds 1, it outperforms the rectangular reference pulse with maximum amplitude and if it falls below 0, the analyzed pulse performs worse than the provided guess pulse. 
The gain is a valuable quantity to ensure a fair comparison between the different analysis methods, regardless of their absolute value at any given point. 
If the absolute value of the FoM would change over several days due to thermal shifts or mechanical disturbances, the calculated gain remains constant.

\subsection{Optimization Workflow}
\label{subsec:workflow}

\begin{figure}
    \centering
    \includegraphics[width=0.48\textwidth]{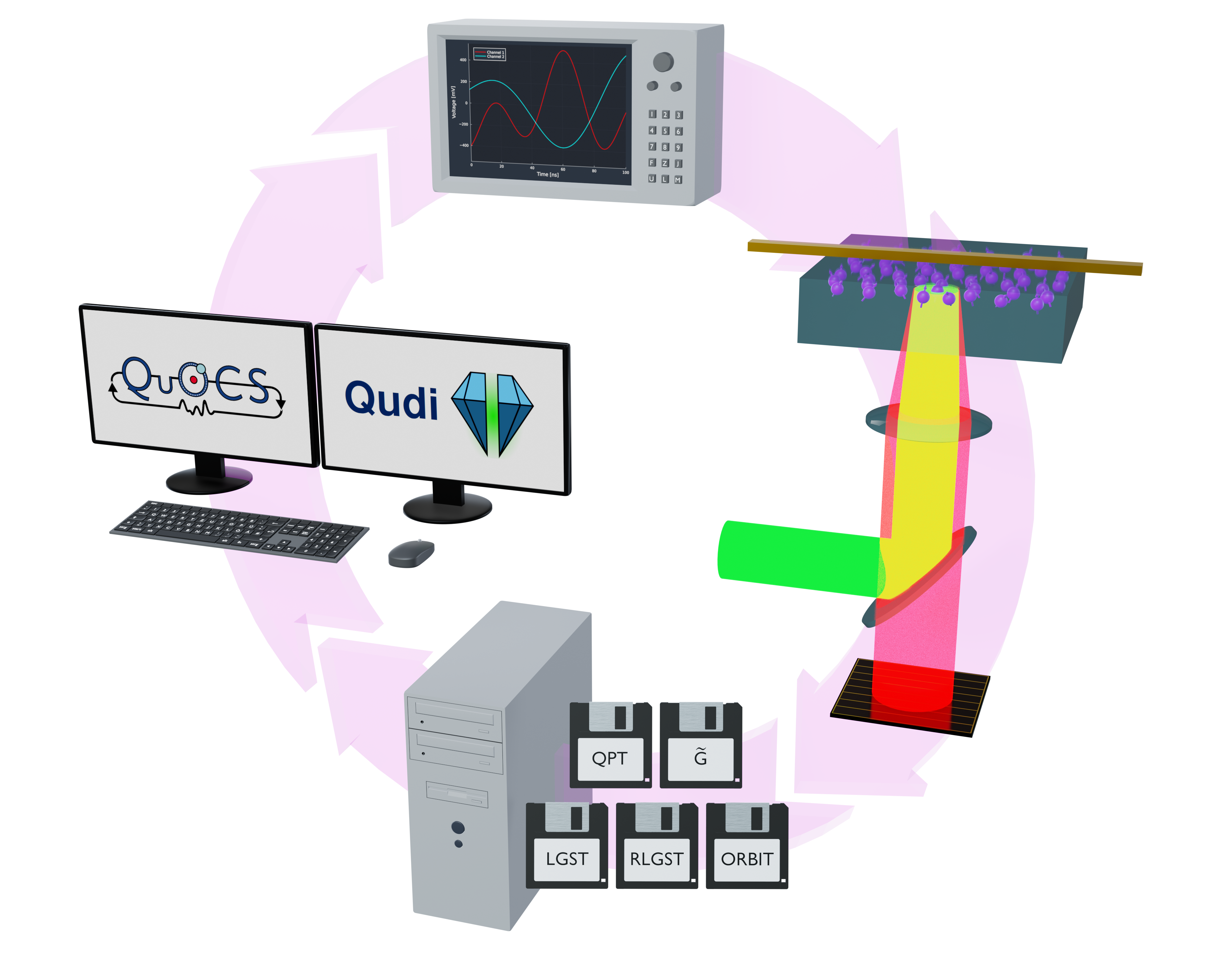}
    \caption{Workflow of the closed-loop QOC experiments. 
    First, we select the analysis method, e.g. RLGST.
    Qudi then constructs the measurement sequence using the optimized pulse shape provided by QuOCS.
    The sequence is uploaded to the AWG to create the microwave signals.
    Finally, a green laser pulse is used to read out the \nv centers and the red fluorescence light is collected to calculate the corresponding FoM.
    The FoM is send to QuOCS which in return computes a new pulse shape and the cycle repeats.}
    \label{fig:workflow}
\end{figure}

The pulse shape of the microwave pulses
\begin{equation}
    \Omega\left(t\right) = a_{x}(t)\cos\left(\omega t\right) + a_{y}(t)\sin\left(\omega t\right)
    \label{eq:pulse_shape}
\end{equation}
can be expressed by the time-dependent amplitudes $a_{x}\left(t\right)$ and $a_{y}\left(t\right)$, which represent the $S_x$ and $S_y$ ($x$- and $y$- spin operator) component of the corresponding gate.
During our closed-loop QOC experiments we vary those time-dependent amplitudes according to the dCRAB algorithm~\cite{rach2015dressing, frank2017autonomous, rembold2020introduction, mueller2022chopped} (see Methods section~\ref{subsec:dcrab_settings}). 

Fig.~\ref{fig:workflow} describes the workflow of such a closed-loop optimization. 
First, we select an analysis method, e.g. RLGST, which defines how the measurement sequence looks like and how the FoM is calculated. 
At the start of the optimization QuOCS~\cite{rossignolo2023quocs} hands Qudi the initial guess pulse. 
The pulse is incorporated in the measurement sequence constructed by Qudi~\cite{binder2017qudi}, our experimental control and data processing software.
The full measurement sequence is uploaded via Qudi to the arbitrary waveform generator (AWG) (Keysight M8195A) which generates the analog waveform and all required digital trigger signals.
The analog waveform signal is sent through a 30~W amplifier (AR-30S1G6) at 80\% gain and then through a straight gold microwave structure on the diamond's surface. 
The \nv centers are initialized and read out via a 532~nm green laser (Novanta Photonics gem 532) and their red fluorescence light is collected by a silicon photo-multiplier (Ketek PE3315-WB-TIA-SP). 
We use a digitizer (Spectrum Instrumentation M4i.4420-x8) to record the fluorescence signal which in turn is analyzed and evaluated by Qudi to obtain the \nv centers' spin state population.
Once finished, Qudi calculates the FoM according to the selected analysis method, while treating the \nv ensemble as one effective qubit. 
The FoM is handed to QuOCS which in turn calculates a new pulse shape and the whole cycle repeats.\\
Following the closed-loop QOC experiment, the optimized pulse is benchmarked via all evaluation methods as well as a complete RB experiment.

\subsection{Cross-Comparison}
\label{subsec:opt_cross_comparison}

\begin{figure*}
    \centering
    \includegraphics[width=0.99\textwidth]{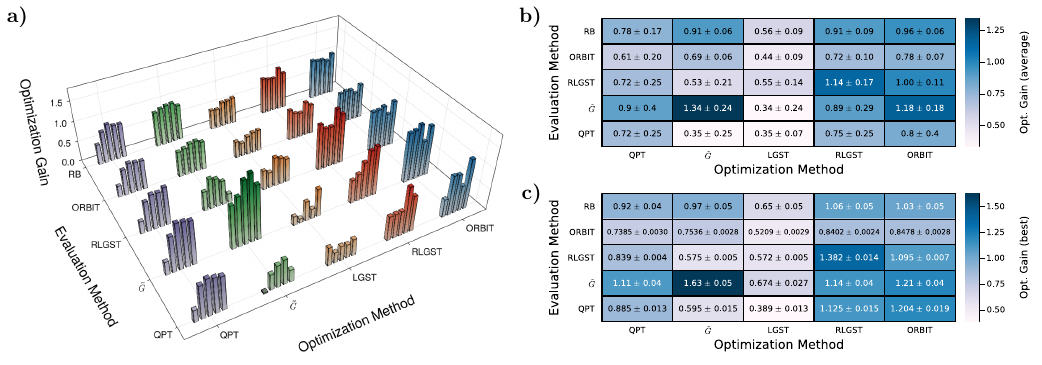}
    \caption{Cross-comparison of the closed-loop QOC experiments.
    \textbf{a)} The x-axis shows the method selected to calculate the FoM during an optimization.
    We perform five optimization runs per method, represented by five bars, whose color denotes the optimization method. 
    Their height corresponds to the achieved gain for the chosen evaluation method displayed in the y-axis and is is the average of $20$ individual measurements.
    \textbf{b)} Average gain of the five optimizations per method together with the standard deviation.
    The abscissa corresponds to the method selected for the FoM calculation during the optimization and the ordinate to the chosen evaluation method to calculate the gain.
    \textbf{c)} The best performing pulse, when evaluated by its optimization method, is re-measured for all other evaluation methods. 
    The error corresponds to the uncertainty of 20 measurements.}
    \label{fig:evaluation}
\end{figure*}

Each optimization setting is repeated five times to investigate the stability and reproducibility of the optimization. 
The cross-evaluation of each experiment is shown in Fig.~\ref{fig:evaluation} where a) displays an overview of all obtained gain values after optimization.
Each column represents the method used for the FoM calculation during the closed-loop QOC experiment, each row represents the evaluation method and the height of the bars corresponds to the achieved gain, Eq.~\eqref{eq:optimization_gain}.
For the evaluation with randomized benchmarking we use the extracted average error rate~\cite{magesan2011scalable, magesan2012characterizing} instead of a FoM to calculate the corresponding gain. 
For a better comparison, we also show the average gain over all five optimizations for each method with the corresponding standard deviation in Fig.~\ref{fig:evaluation}~b).
In Fig.~\ref{fig:evaluation}~c) we show the gains of the pulse that performs best upon evaluation with its own optimization method for each of the five runs.

%
%
%
%
\subsection{Optimization with QPT}

For QPT, we observe a homogeneous gain progression across the whole column, i.e. for all evaluation methods. 
Evaluated with QPT, the achieved gain is on the same level as when evaluated by RLGST and RB. 
The first optimization run did not find a very good solution, which can happen because of the limited settings for the optimizer and can safely be marked as an outlier. 
Note, however, that this outlier leads to a large variance of the average values in Fig.~\ref{fig:evaluation}~b). 
%
%


%
%
%
%
\subsection{Optimization with $\Tilde{G}$}

Optimizing the pulse shape with $\Tilde{G}$ leads to an outstanding improvement of the optimized pulse over the reference pulse when being evaluated with $\Tilde{G}$ itself, resulting in an average gain of $1.34$ and the highest achieved value of all measurements of $1.63$.
However, the large improvement in the gain compared to the reference pulse is not retained for the other evaluation methods.
We obtain low performances for QPT and RLGST but high gain values if evaluated with ORBIT and RB. 
Despite their similarities, a large improvement in ORBIT must not translate to a large improvement in RLGST and vice versa at the selected circuit length. 
Especially for the evaluation with RLGST, while still showing an improvement, the optimization with $\Tilde{G}$ shows the lowest overall achieved gain for all optimization methods. 
The same applies to the evaluation with QPT together with the LGST-optimized pulses. 
While the measurements for $\Tilde{G}$ and QPT consist of combinations of three gates and are therefore sensitive to errors on the same time-scale, $\Tilde{G}$ evaluates the performance of the full gate-set and QPT only the one of $G_2$. 
$\Tilde{G}$ tries to maximize the overlap of predefined expectation values according to Eq.~\eqref{eq:fom_gtilde}. 
Thus, $\text{FoM}_{\Tilde{G}}$ adjusts the pulse shape of $G_2$ during the optimization so that it is optimized for these expectation values and compensates for the errors of the other pulses in the gate-set, Eq.~\eqref{eq:gate_set}. 
While this leads to an overall improvement of the gate-set's performance, the fidelity of $G_2$ alone is not maximized, leading to a reduced gain in QPT. 


%
%
%
%
\subsection{Optimization with LGST}

Next, we analyze the optimizations with LGST via $\text{FoM}_\text{LGST}$ defined by Eq.~\eqref{eq:fom_lgst}. 
As expected, we observe the lowest gain for all evaluation methods across all optimization methods.
Still, we always observe a positive gain, i.e., QuOCS is able to improve the gate-set's performance by varying the pulse shape of $G_2$.
A major problem for optimization with LGST is that it is not able to differentiate between the reference and guess pulse as shown in Fig.~\ref{fig:Figure_1}~f) and might therefore also misinterpret good QOC pulses. 
Additionally, small variations in the pulse shape and measurement errors can lead to an enormous change of the FoM due to the calculation of the estimates via the Gram matrix in Eq.~\eqref{eq:lgst_G_hat}.
The inversion of a matrix filled with measured, and therefore noisy, values as well as constantly changing entries due to the variations during the dCRAB search, is propagating and amplifying small changes through the whole derivation~\cite{lefebvre2000propagation}. 
This results in a strongly fluctuating FoM during the optimization (see supplementary information~\cite{supplementary}).
QuOCS can deal with such experimental downsides but it still influences the search for an optimal solution negatively.
%


%
%
%
%
\subsection{Optimization with RLGST}

Optimizing the pulse shape via RLGST shows a large average gain for all methods. 
We are able to find an optimized gate-set that reaches gains exceeding a value of $1$ for almost all evaluation methods, as shown by Fig.~\ref{fig:evaluation}~c). 
%
%
The average value of $0.75$ in Fig.~\ref{fig:evaluation}~b) and the highest value with $1.13$ in Fig.~\ref{fig:evaluation}~c) when being evaluated with QPT even exceeds the QPT optimization itself. 
%
%
Therefore, not only the individual gate performance greatly increases, but also that of the entire gate-set, which is reflected by the high gain at different time-scales for ORBIT and $\Tilde{G}$. 
Nevertheless, QPT and $\Tilde{G}$ show a large variance in the obtained gain relative to the mean value, as RLGST considers the performance for longer circuits.
Thus, not every optimized pulse shape must necessarily work equally well for a small number of applied pulses.


%
%
%
%
\subsection{Optimization with ORBIT}

The optimization with ORBIT shows a large improvement for all evaluation methods. 
If we take the average over all evaluation methods, ORBIT provides the best performance. 
Through an optimization with ORBIT, we find gate-sets that perform universally well, regardless of the used evaluation method. 
For RLGST and $\Tilde{G}$ we observe large improvements of the gain, reaching and even surpassing the performance of the reference pulse. 
The method also shows the highest average gain of all methods when evaluated with RB, which is only just below the reference pulse. 
Upon evaluation with QPT the optimized pulses on average outperform the ones achieved when being optimized via QPT itself. 
We observe a large variance between the different optimizations for QPT evaluation, while all other methods exhibit a small variance relative to their mean value. 
Like for the optimization with RLGST, this can be explained by the fact that ORBIT optimizes the full gate-set's performance while QPT evaluates the single pulse $G_2$. 
Thus, the performance of an individual gate must not be maximized.

\subsection{Circuit length dependence}
\label{subsec:L_dep}

\begin{figure}
    \centering
    \includegraphics[width=0.48\textwidth]{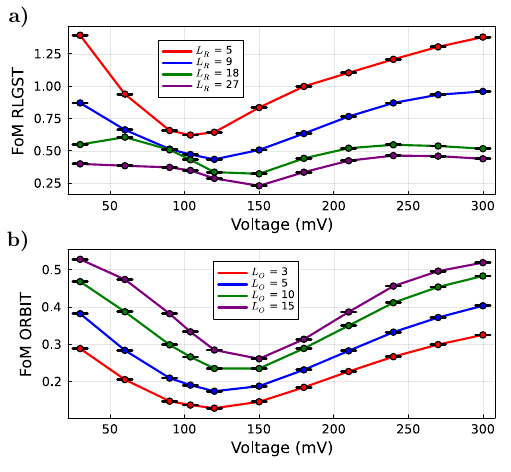}
    \caption{Dependence of the FoMs on the circuit length.
    Linear sweep of the guess pulse amplitude for different circuit lengths $L$ for \textbf{a)} RLGST and \textbf{b)} ORBIT.
    The y-axis denotes the achieved FoM.
    Each data point represents the mean value of $20$ measurements with the corresponding uncertainty and the chosen circuit length is visualized by the different colors.}
    \label{fig:4}
\end{figure}

The methods RLGST and ORBIT allow to easily probe how the system's dynamics at different time-scales affect the FoM by varying the selected circuit length. 
For this reason, we first examine how our defined FoMs for RLGST and ORBIT are affected by a linear change of the pulse amplitude for different $L$. 
The circuit length differs for the two methods: for RLGST it indicates the number of applied gates from the gate-set Eq.~\eqref{eq:gate_set}, while for ORBIT it stands for the number of applied Clifford gates. 
On average, one Clifford gate consists of $\approx 1.8$ gates from our gate-set Eq.~\eqref{eq:gate_set} and thus we choose the gate-string length of RLGST to match the average number of gates in ORBIT. 
Again we average over 300 random gate-strings per method. 
The measurement results are shown in Fig.~\ref{fig:4}~a) for RLGST and in Fig.~\ref{fig:4}~b) for ORBIT. 
We observe a shift of the minimum FoM, i.e. the best performing pulse amplitude, to larger amplitudes for increasing circuit length for both methods. 
This correlation can be explained by the strong beating observed in the Rabi experiment in Fig.~\ref{fig:Figure_1}~a). 
Hopping from peak to peak of the Rabi oscillation is analogous to the application of a series of $\pi$-pulses on the system. 
If the $\pi$-pulse has a fixed pulse length, the amplitude of the pulse must be increased for subsequent applications to compensate for the shift in the Rabi frequency caused by the beating. 
Thus, to find the best performing pulse for increasing number of repeated applications, a.k.a. $L$, the amplitude, in the form of a voltage in our case, needs to be adjusted upwards. 

For ORBIT, the FoM gets worse (increases) with increasing circuit length as errors accumulate. 
This is expected as ORBIT evaluates the gate-set's performance at a fixed circuit length of the full RB curve, which exhibits an exponential decay with $L$. 
Therefore, increasing $L$ samples the RB curve at a point of lower survival probability corresponding to a higher FoM per our definition. 
For RLGST we observe the exact opposite behaviour, where an increased circuit length leads to a lower FoM. 
The FoM of RLGST in Eq.~\eqref{eq:fom_rlgst} in the Methods is based on the estimated error matrices for the initial state, POVM and the full gate-set, which are only accurate within the linear assumption~\cite{gu2021randomized}. 
Beyond this regime, the prediction error diverges~\cite{gu2021randomized} and the FoM decreases as the error matrices are under-estimated. 
Nonetheless, the argument of suitability of the FoM for our closed-loop QOC experiments holds due to the distinct minimum and the irrelevance of its absolute value as long as it guides the optimizer to a better solution. 
Importantly, both FoMs agree on the optimal amplitude for comparable $L$ (minima of the curves with the same color code).

\subsection{Optimizations with varying circuit length}
\label{subsec:opt_var_L}

\begin{figure*}
    \centering
    \includegraphics[width=0.99\textwidth]{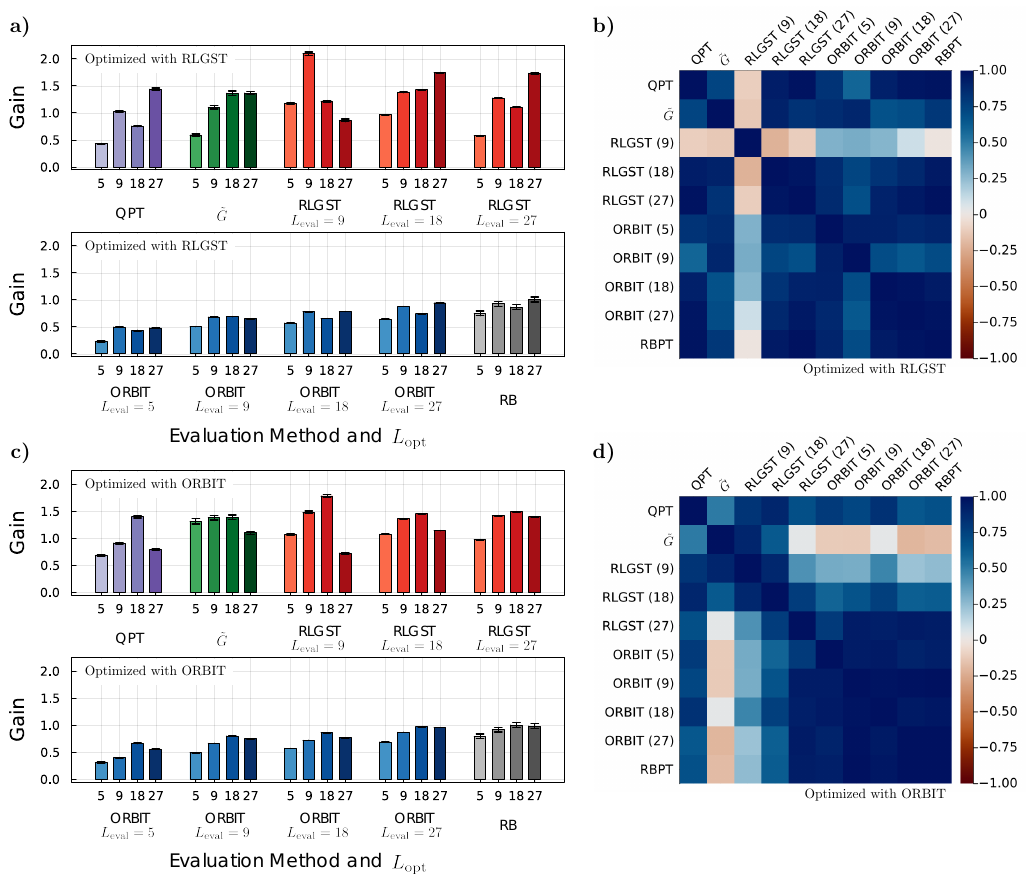}
    \caption{Influence of the circuit length on the optimization results.
    \textbf{a)} Optimization runs with RLGST for four different circuit lengths $L=5,9,18,27$, represented by the four bars.
    The ordinate shows the gain achieved per optimization for the different evaluation methods displayed in the abscissa as an average of $20$ measurements and the error bars denote the corresponding uncertainty. 
    $L_{eval}$ indicates the selected circuit length for the evaluation with RLGST and ORBIT.
    \textbf{b)} Correlation matrix of the RLGST optimizations.
    The color-scale denotes the correlation between the different methods, i.e. if we observe the same dependence of the achieved gain on the circuit length.
    \textbf{c)} Gain of the optimization runs for different circuit lengths with ORBIT. 
    %
    %
    \textbf{d)} Correlation matrix of the ORBIT optimizations.}
    \label{fig:5}
\end{figure*}

Next, we investigate how a change of the circuit length affects the optimization result. 
We therefore repeat the optimizations for RLGST and ORBIT once for different circuit lengths $L$. 
The results are shown in Fig.~\ref{fig:5}~a) for RLGST and Fig.~\ref{fig:5}~c) for ORBIT. 
We again choose the circuit length of RLGST such that it matches the average number of Clifford gates of ORBIT. 
For better readability we only display the average number of gates for ORBIT in Fig.~\ref{fig:5}~c). 
The abscissa shows the chosen evaluation method, e.g. RLGST with a circuit length of $L_{\text{eval}}=18$, the ordinate shows the achieved gain.
The four bars per evaluation method represent the individual optimizations for a chosen circuit length.
We omit the evaluation with $L_{\text{eval}}=5$ for RLGST as the difference between the reference pulse and the guess pulse is within our measurement error and we are therefore not able to calculate the corresponding gain.
Since we only perform one optimization per method and $L_{\text{opt}}$, we cannot exclude the possibility that some optimizations under- or over-performed due to the probabilistic nature of the optimization (caused by the measurement noise and the optimization algorithm). 
However, because of the small variance between different runs in Fig.~\ref{fig:evaluation}, no significant change between repeated optimizations for the same $L_{\text{opt}}$ is expected.
%
%

%
For both optimization methods the overall achieved gain evaluated by ORBIT increases with increasing gate-string length $L_{\text{eval}}$.
As the guess pulse is significantly longer than the reference pulse, the difference in their FoMs increases with the circuit length due to, inter alia, decoherence. 
This increases the distance between the FoMs of the reference and guess pulse.
If the optimized pulse can compensate for this decoherence we observe an overall higher gain with increasing $L_{\text{eval}}$ for the evaluation with ORBIT. 
%
%

Fig.~\ref{fig:5}~b) and d) show the correlation matrices for the optimizations with different circuit lengths presented in Fig.~\ref{fig:5}~a) and c) for RLGST and ORBIT, respectively. 
The correlation, described in the Methods~\ref{subsec:corr_mat}, looks at the behavior between the gains of different optimized pulses and compares it to other evaluation methods. 
E.g., if we observe that an increase of the circuit length leads to an increase of the gain when evaluated with ORBIT at $L=27$, a high correlation with RB tells us that the method shows the same tendency. 

When optimizing the gate-set via RLGST, we observe a high correlation between all evaluation methods except for the evaluation with RLGST at $L_{\text{eval}}=9$, as shown by the correlation matrix in Fig.~\ref{fig:5}~b).
An increase of the circuit length $L_{\text{opt}}$ leads to an increase of the achieved gain. 
At large $L_{\text{opt}}$, small gate errors accumulate and cause a stronger change of the FoM, leading to a more precise and overall better optimization. 
This is also reflected by the absolute value of the gain, which for large $L_{\text{opt}}$ almost always exceeds $1$, when evaluated with RLGST, $\Tilde{G}$ and QPT. 
For RLGST at $L_{\text{eval}}=9$ the gain even exceeds a value of $2$, when optimized with the same circuit length. 
As the evaluation at this circuit length poorly correlates with all other evaluations, it can be assumed that the system exhibits certain dynamics exclusive to this time-scale and chosen measure. 
%
%

%
The optimization with ORBIT at different circuit lengths shows again a high correlation between QPT, RB and the different ORBIT evaluations. 
The correlation matrix for ORBIT is shown in Fig.~\ref{fig:5}~d).
When evaluating the pulse with RLGST, we observe an increase of the correlation with ORBIT and RB for increasing $L_{\text{eval}}$.
For $L_{\text{eval}}=27$, the observed dependence of the gain on the circuit length correlates very well with that of ORBIT.
At large circuit lengths, the FoM's of RLGST and ORBIT are dominated by the same dynamics and thus converge to similar solutions, i.e. similar performing gate-sets.
We observe again, that an increase of $L_{\text{opt}}$ leads to a higher gain.
Through the high correlation between the methods we can conclude that the optimization with ORBIT at $L_{\text{opt}}=27$ significantly under-performed and a higher gain is expected for repeated optimizations. 
Although showing high gains significantly exceeding $1$, the evaluation with $\Tilde{G}$ shows the smallest correlation with the other methods.
Different circuit lengths during the optimization with ORBIT have thus no significant effect on the gain observed in $\Tilde{G}$.
%



\section{Discussion}
\label{sec:discussion}

Based on our detailed investigation, we can now discuss to what extent the individual methods are suitable for closed-loop optimizations with spin ensembles treated as one effective qubit or with qubits in general.
We would like to emphasize that all methods have evidently been able to improve the performance of our gate-set. 

Out of all investigated methods LGST achieves the lowest gains. 
Measurement errors dominate the provided estimates and therefore the calculated FoMs.
As a result, the optimizer sees no clear change for pulses that perform differently well, leading to an overall low optimization gain.
If the FoM is calculated using solely the expectation values of LGST, i.e. $\Tilde{G}$, significantly larger gains can be achieved.
Since optimizations via $\Tilde{G}$ maximize the overlap of the predefined expectation values to their target and allow to simply add additional gates to the gate-set, the method is particularly suitable if the gate-set is to be optimized for specific applications of the gates to arbitrary basis states.
The method is heavily biased towards optimizing for its specific measurement sequence and the associated average circuit length.
This leads to a high gain when evaluated by the $\Tilde{G}$ FoM itself but must not perform equally well when evaluated by other methods and on different time-scales or with other circuits.

QPT, unlike all the other methods, does not take SPAM errors into account.
For larger SPAM errors, the correlation to other methods is expected to diminish as does its utility. 
Such is the case for our system, where an optimization based on QPT never beats the gate-set with the reference pulse.
The optimization process is furthermore limited to one single gate at a time. 
While the optimization of an entire gate-set might be possible through iterative optimizations, the other methods offer a much better and clearer approach. 
Despite this, our FoM based on QPT is able to optimize the gate-set such that it shows improvements compared to the guess pulse for all investigated methods. 
Among all methods, QPT requires the fewest measurements, which can be beneficial when working with systems with low signal-to-noise ratio. 
The overall best results are achieved by ORBIT. 
Not only does ORBIT show universally high gains across all methods, but the optimized gate-sets also outperform those of other optimization methods when evaluated with their own FoM, regardless of their time-scale. 
This means that the ORBIT-optimized pulse drastically increases the performance of the entire gate-set and can simultaneously improve the single-gate fidelity of the selected gate. 
As shown in the previous section, long gate-string lengths are recommended for the optimization as errors accumulate, leading to larger changes in the defined FoM and therefore a more precise optimization. 
Too large $L$ are not favourable, though, since information gets lost in noise and accumulated errors. 
When increasing the circuit length, we expect to find a sweet spot for the optimization where such dynamics are captured and weighed optimally. 
In our case we are limited by the fluence (see supplementary information~\cite{supplementary}) and therefore heating induced by long circuits. 

One downside of ORBIT is that randomized benchmarking usually requires the use of Clifford gates which are generated by the over-full gate-set, Eq.~\eqref{eq:gate_set}. 
Additional gates that are to be optimized cannot simply be appended to the gate-set but need to be interleaved into the Clifford gates if possible. 

This problem can largely be circumvented by the optimization via RLGST. 
Additional gates can simply be added to the gate-set. 
%
%
Optimizations via RLGST achieve similar high gains to ORBIT across all methods, i.e. are able to create a universally well-performing gate-set.
Analogous to ORBIT, large circuit lengths are preferred when working with RLGST. 
At large circuit lengths, RLGST evaluates the system dynamics similar to ORBIT making the method ideal when working with a overfull gate-set as in our case. 
We show that the method can be used far beyond its intended linear (small error) working regime~\cite{gu2021randomized} to define a suitable FoM for closed-loop experiments. 
Moreover, while not directly investigated in our work, the method also allows to work with an incomplete gate-set~\cite{gu2021randomized}, which can be of use for experimental systems with limited control. 
In addition, the method provides estimates of the gate-set and states which can be accessed during the optimization to obtain deeper insights into the system dynamics~\cite{wittler2021integrated}. 

%
%
%
%

In conclusion, through the definition of our FoMs based on RLGST and ORBIT we are able to create a gate-set which performs universally well across all investigated evaluation methods, regardless of their time-scale.
We provide a detailed comparison of several gate and gate-set analysis methods and how these can be applied in a closed-loop QOC experiment, tested with a macroscopic ensemble of \nv centers.
Since the occurring types of error are not exclusive to our system but exist in a wide variety of systems, our work can act as a useful handbook for experimentalists trying to overcome similar problems via QOC. 
The methods and the corresponding FoMs that work well for our system are expected to perform similar for other systems with comparable dynamics~\cite{scholten2021widefield, gottscholl2021spin, gong2023coherent, castelletto2020silicon} and will also be applicable to systems with significantly smaller errors~\cite{bartling2024universal, werninghaus2021leakage, dehollain2016optimization}.

%


\section{Methods}
\label{sec:methods}

\subsection{Definitions of the Figures of Merit}
\label{subsec:fom_definitions}

In addition to our gate-set Eq.~\eqref{eq:gate_set}, we define the SPAM-set to be
\begin{equation}
    \begin{split}
        \mathcal{F}&=\lbrace F_0, F_1, F_2, F_3 \rbrace\\
    &= \lbrace G_0, G_1, G_3, G_5 \rbrace \; .
    \end{split}
    \label{eq:spam_gate_set}
\end{equation}
%

%
%

Starting with QPT, we measure the expectation values
\begin{equation}
    p_{ij} = \langle\langle E | F_i G_2 F_j | \rho\rangle\rangle
    \label{eq:probabilities_values_qpt}
\end{equation}
for $F_{i,j} \in \mathcal{F}$ to evaluate the performance of $G_2$ along all axes and reconstruct the final states $\vert 0 \rangle$, $\vert -1 \rangle$, $\vert + \rangle = 1/\sqrt{2} \left( \vert 0 \rangle + \vert -1 \rangle \right)$ and $\vert - \rangle = 1/\sqrt{2} \left( \vert 0 \rangle + i \vert -1 \rangle \right)$ after application of the gate. 
Finally, we calculate the process matrix $\chi$ according to~\cite{frank2017autonomous, nielsen2002quantum}, which defines the full linear map of our gate $G_2$.
The FoM is then defined as the Frobenius norm of the distance between the measured and the target process matrix $\chi_{\text{T}}$ for the perfect gate:
\begin{equation}
    \text{FoM}_{\text{QPT}} = \sqrt{\text{tr}\left(\left(\chi-\chi_{\text{T}}\right)^\dagger\left(\chi-\chi_{\text{T}}\right)\right)} \; .
    \label{eq:fom_qpt}
\end{equation}
%

%
%

In contrast to QPT, quantum gate-set tomography~\cite{blume2013robust, merkel2013self} takes SPAM errors directly into account by measuring the expectation values of Eq.~\eqref{eq:probabilities_values_qpt} for all $G_k \in \mathcal{G}$:
\begin{equation}
\begin{split}
    p_{ijk} &= \langle\langle E | F_i G_k F_j | \rho\rangle\rangle \\
    &= \left( \widetilde{G}_k \right)_{ij}.
    \end{split}
    \label{eq:lgst_probabilities}
\end{equation}
Following Refs.~\cite{blume2013robust, greenbaum2015introduction}, assuming a perfect idle gate $\mathds{1}$ leads to the estimates
\begin{equation}
    \begin{split}
        \hat{G}_k &= \widetilde{G}_0^{-1} \widetilde{G}_k, \\
        \vert \hat{\rho} \rangle\rangle &= \widetilde{G}_0^{-1} \vert \widetilde{\rho}\rangle\rangle\\
        \vert \hat{E} \rangle \rangle &= \vert \widetilde{E} \rangle \rangle \, .
    \end{split}
    \label{eq:lgst_G_hat}
\end{equation}
using the inverse Gram matrix $\widetilde{G}_0^{-1}$ to redistribute the errors to all other gates. 
As the expectation values~\eqref{eq:lgst_probabilities} are gauge invariant, the estimates~\eqref{eq:lgst_G_hat} need to be gauge-corrected. 
Following the derivations provided in Refs.~\cite{blume2013robust, greenbaum2015introduction}, we then obtain the gauge-transformed LGST estimates $\hat{G}^*_k,~\vert \hat{\rho}^* \rangle\rangle$ and $\vert \hat{E}^* \rangle\rangle$. 
Usually, a maximum likelihood estimation (MLE) is used next to find the closest physically correct version of the obtained estimates, i.e., a $\hat{G}^*_k$ that corresponds to a CPTP map. 
However, such an MLE is computationally demanding, making it unsuitable for our closed-loop QOC experiments. 
We therefore use the estimates provided by LGST to evaluate the performance of our gates $G_k$ through the difference to their target $T_k$: 
\begin{equation}
    \text{FoM}_\text{LGST} = \sqrt{ \sum_{k=1}^{K+1} \text{tr}\left(\left(\hat{G}^*_k-T_k\right)^\dagger\left(\hat{G}^*_k-T_k\right)\right)} \,
    \label{eq:fom_lgst}
\end{equation}
with $\hat{G}^*_{K+1}\equiv \vert\hat{\rho}^*\rangle\rangle\langle\langle \hat{E}^*\vert$.
We sum up the squared norms for each gate difference similar to the sum of squared residuals in the least-squares method~\cite{van1991the}. 
Additionally, we define another FoM 
\begin{equation}
    \text{FoM}_{\Tilde{G}} = \sqrt{ \sum_{k=1}^{K} \text{tr}\left(\left(\Tilde{G}_k-\Tilde{T}_k\right)^\dagger\left(\Tilde{G}_k-\Tilde{T}_k\right)\right)} \, ,
    \label{eq:fom_gtilde}
\end{equation}
based on the matrices $\Tilde{G}_k$ of Eq.~\eqref{eq:lgst_probabilities} filled with the measured expectation values and their difference to the corresponding target values $\Tilde{T}_k$. 
While not providing an estimate, these expectation values are gauge invariant by definition. 

%
%

Another method which takes SPAM errors into account is randomized benchmarking~\cite{knill2008randomized, magesan2011scalable, magesan2012characterizing}.
Randomized benchmarking uses Clifford gates, which are composed of gates from our gate-set Eq.~\eqref{eq:gate_set}~\cite{epstein2014investigating}, to create multiple random circuits of length $L$, where the final gate flips the spin back to its initial state.
The average recovered population, the so-called survival probability $p_{s}$, decays exponentially with the number of applied Clifford gates~\cite{magesan2011scalable} and the average error per gate can be extracted from the decay parameter.
ORBIT then allows to optimize the fidelity of the applied gates by increasing the survival probability at an arbitrary gate-string length $L_{\text{opt}}$~\cite{kelly2014optimal}.
Therefore, we define the corresponding FoM for our closed-loop QOC experiments with ORBIT by
\begin{equation}
    \text{FoM}_\text{ORBIT} = 1 - p_{s}(L_{\text{opt}}) \; .
    \label{eq:fom_orbit}
\end{equation}

%
%

RLGST~\cite{gu2021randomized} combines the idea of gate-set tomography and randomized circuits to obtain an estimate of the gate-set with little computational overhead within a linear approximation.
We start by measuring the expectation values
\begin{equation}
    p_i = \langle\langle E | \mathcal{C}_i | \rho \rangle\rangle
    \label{eq:rlgst_probabilities}
\end{equation}
for $N$ randomly chosen circuits $\mathcal{C}_i$ of length $L$ with $i=1...N$.
Following the calculations provided in Ref.~\cite{gu2021randomized}, the expectation values of Eq.~\eqref{eq:rlgst_probabilities} are then used to obtain an estimate for the error matrices $e_j$, which describe how the measurement deviates from the expected target:
\begin{equation}
    \begin{split}
        \hat{G}_k &= \left(\mathds{1}+e_k\right) \cdot T_k,\\
        \vert \hat{\rho} \rangle \rangle &= \left(\mathds{1}+e_\rho\right)\cdot \vert \rho_T\rangle\rangle,\\
        \vert \hat{E} \rangle \rangle &= \left(\mathds{1}+e_E\right) \cdot \vert E_T \rangle\rangle \, . 
    \end{split}
    \label{eq:rlgst_definitions}
\end{equation}
Since we only vary a single gate during our optimizations, we expect to be reasonably close to the actual gauge for the estimates and defining the FoM for RLGST equivalent to Eq.~\eqref{eq:fom_lgst} we obtain
\begin{equation}
    \text{FoM}_\text{RLGST} = \sqrt{ \sum_j \text{tr}\left(e_j^\dagger e_j\right)}.
    \label{eq:fom_rlgst}
\end{equation}

\subsection{Correlation Matrix}
\label{subsec:corr_mat}

The correlation matrices are calculated according to Ref.~\cite{helwig2017data} via
\begin{equation}
    \label{eq:corr_mat}
    M = \begin{pmatrix} 
        1 & r_{12} & r_{13} & \dots & r_{1p} \\
        r_{21} & 1 & r_{23} & \dots & r_{2p} \\
        r_{31} & r_{32} & 1 & \dots & r_{3p} \\
        \vdots & \vdots & \vdots & \ddots & \vdots \\
        r_{p1} & r_{p2} & r_{p3} & \hdots & 1 \\
    \end{pmatrix}
\end{equation}

with the correlation coefficient
\begin{equation}
    \label{eq:pearson}
    r_{jk} = \frac{\sum_{i=1}^{n} \left( x_{ij} - \bar{x}_j \right) \left( x_{ik} - \bar{x}_k \right)}{\sqrt{\sum_{i=1}^{n} \left( x_{ij} - \bar{x}_j \right)^2} \sqrt{\sum_{i=1}^{n} \left( x_{ik} - \bar{x}_k \right)^2}}
\end{equation}

where the bar indicates the average over the indexed column of the data matrix
\begin{equation}
    \label{eq:data_mat}
    X = \begin{pmatrix} 
        x_{11} & x_{12} & \dots & x_{1p} \\
        x_{21} & x_{22} & \dots & x_{2p} \\
        \vdots & \vdots & \ddots & \vdots \\
        x_{n1} & x_{n2} & \hdots & x_{np} \\
    \end{pmatrix} \; .
\end{equation}

The rows ($n$) of $X$ are the four optimizations with varying $L$, i.e. the four pulses that are compared, and the columns ($p$) describe the different evaluation methods used.

\subsection{dCRAB Settings}
\label{subsec:dcrab_settings}

The dCRAB method expands updates to the amplitudes in a randomly selected sub-set of functions sampled from a ``chopped'' basis. 
The frequencies for the update pulses are randomly selected from the interval $\left[ 0, \frac{ 2 \pi \cdot n }{T} \right]$ where we choose $n=4$ as the maximum allowed number of oscillations of the pulse envelope over the course of the pulse duration $T$. 
Several super-iterations with newly randomized basis vectors restart the search process by adding new search directions to avoid getting trapped in local minima. 
In our case we perform 3 super-iterations. 
We use the Nelder-Mead search method for the expansion parameters and select two basis vectors per pulse per super-iteration for the basis expansion. 
In this way, the search parameters for the optimization can be kept to a minimum, to ensure a quick convergence. 
If a potential improvement of the FoM does not exceed the measurement error, QuOCS will re-evaluate the evaluation step up to three times to ensure an accurate interpretation of the parameter landscape~\cite{rossignolo2023quocs}.
We determine the standard deviation by measuring the FoM for the guess pulse 100 times. 
A super-iteration stops, if the improvement of the FoM within the last 200 evaluation steps does not exceed the standard deviation.
To account for experimental drifts in the FoM, the current best pulse is re-measured every 30 minutes.

\subsection{Data Normalization and Measurement Sequence}
\label{subsec:data_normalization}

\begin{figure}
    \centering
    \includegraphics[width=0.48\textwidth]{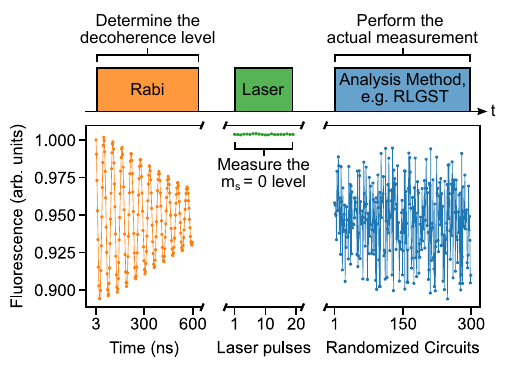}\\
    \caption{The measurement sequence for any experiment consists of three parts: a Rabi measurement to determine the decoherence level, a measurement of the $\vert 0\rangle\rangle$ fluorescence level and the actual measurement sequence, e.g. RLGST, which is then converted to expectation values by the previously obtained fluorescence levels.}
    \label{fig:measurement_sequence}
\end{figure}
For any performed experiment, the measurement sequence consists of three parts which are measured simultaneously, as depicted in Fig.~\ref{fig:measurement_sequence}. 
We start with a Rabi measurement at maximum amplitude for 600~ns to determine the decoherence level of our system and afterwards measure a series of 20 laser pulses to obtain the fluorescence level of the $\vert 0\rangle\rangle$ state. 
Those levels are used to convert the fluorescence signal of the selected analysis method, to expectation values. 
Possible changes of the measurement contrast, through e.g. laser power fluctuations, are thus taken into account by adjusting the measured expectation values such that $0$ and $1$ still correspond to the minimum and maximum achievable fluorescence.


\section{Data Availability}
\label{sec:data_availability}

The data presented in this study is available from the corresponding authors on reasonable request.


\newpage
\bibliography{references}


\section{Acknowledgements}
We thank Ressa Said, Timo Joas, Marco Rossignolo and Maximilian Kraus for helpful discussions. \\
This work was funded by the German Federal Ministry of Research (BMBF) by future cluster QSENS (No. 03ZK110AB) and projects DE-Brill (No. 13N16207), SPINNING (No. 13N16210 and No. 13N16215), DIAQNOS (No. 13N16463), quNV2.0 (No. 13N16707), QR.X (No. 16KISQ006) and Quamapolis (No. 13N15375), DLR via project QUASIMODO (No. 50WM2170), Deutsche Forschungsgemeinschaft (DFG) via Projects No. 386028944, No. 445243414, and No. 387073854 and Excellence Cluster POLiS (UP 33/1 Projekt-ID: 422053626), Excellence Cluster Matter and Light for Quantum Computing (ML4Q) EXC 2004/1 – 390534, and Helmholtz Validation Fund project “Qruise” (HVF-00096).
This project has also received funding from the European Union’s HORIZON Europe program via projects QuMicro (No. 101046911), SPINUS (No. 101135699), C-QuENS (No. 101135359), QCIRCLE (No. 101059999) and FLORIN (No. 101086142), European Research Council (ERC) via Synergy grant HyperQ (No. 856432) and Carl-Zeiss-Stiftung via the Center of Integrated Quantum Science and technology (IQST) and project Utrasens-Vir (No. P2022-06-007).


\section{Author Contributions}
P.J.V. and T.R. contributed equally to the project.
P.J.V, T.R., M.M.M, and F.M. conceived the experiments. 
P.J.V., T.R. and M.G.H. implemented the closed-loop optimizations. 
P.J.V performed the measurement.
P.J.V. and T.R. derived the figures of merit, analyzed the data and wrote the manuscript.
M.M.M, F.M., T.C. and F.J. supervised the project.
All authors read and commented on the manuscript. 
%




\section{Competing Interests}
The authors declare no competing interests.


\end{document}


\title{Supplementary Information: Gate-set evaluation metrics for closed-loop optimal control on nitrogen-vacancy center ensembles in diamond}

\author{Philipp J. Vetter}
\thanks{These authors contributed equally to this work. \\Corresponding author: philipp.vetter(at)uni-ulm.de}
\affiliation{Institute for Quantum Optics, Ulm University, Albert-Einstein-Allee 11, 89081 Ulm, Germany}
\affiliation{Center for Integrated Quantum Science and Technology (IQST), 89081 Ulm, Germany}
\author{Thomas Reisser}
\thanks{These authors contributed equally to this work. \\Corresponding author: philipp.vetter(at)uni-ulm.de}
\affiliation{Peter Grünberg Institute -- Quantum Control (PGI-8), Forschungszentrum J\"ulich GmbH, D-52425 Germany}
\affiliation{Institute for Theoretical Physics, University of Cologne, D-50937 Germany}
\author{Maximilian G. Hirsch}
\affiliation{Institute for Quantum Optics, Ulm University, Albert-Einstein-Allee 11, 89081 Ulm, Germany}
\affiliation{Center for Integrated Quantum Science and Technology (IQST), 89081 Ulm, Germany}
\affiliation{Current address: NVision Imaging Technologies GmbH, Wolfgang-Paul-Straße 2, 89081 Ulm, Germany}
\author{Tommaso Calarco}
\affiliation{Peter Grünberg Institute -- Quantum Control (PGI-8), Forschungszentrum J\"ulich GmbH, D-52425 Germany}
\affiliation{Institute for Theoretical Physics, University of Cologne, D-50937 Germany}
\affiliation{Dipartimento di Fisica e Astronomia, Università di Bologna, 40127 Bologna, Italy}
\author{Felix Motzoi}
\affiliation{Peter Grünberg Institute -- Quantum Control (PGI-8), Forschungszentrum J\"ulich GmbH, D-52425 Germany}
\affiliation{Institute for Theoretical Physics, University of Cologne, D-50937 Germany}
\author{Fedor Jelezko}
\affiliation{Institute for Quantum Optics, Ulm University, Albert-Einstein-Allee 11, 89081 Ulm, Germany}
\affiliation{Center for Integrated Quantum Science and Technology (IQST), 89081 Ulm, Germany}
\author{Matthias M. Müller}
\affiliation{Peter Grünberg Institute -- Quantum Control (PGI-8), Forschungszentrum J\"ulich GmbH, D-52425 Germany}

\date{March 2024}

\maketitle

\tableofcontents

\section{Hilbert-Schmidt space}
\label{sec:HS_space}

To simplify the notation throughout our manuscript we express our states and operations in the so-called \textit{Hilbert-Schmidt} (HS) space~\cite{kitaev2002classical, greenbaum2015introduction}. \\
For this, we normalize the Pauli matrices by their dimension $P_i \rightarrow P_i / \sqrt{d}$ where $d = 2$ and choose them as our basis
\begin{equation}
\label{eq:rescaled_pauli_basis}
    P_k \in \bigg\{ \frac{\mathbb{1}}{\sqrt{2}}, \frac{\sigma_x}{\sqrt{2}}, \frac{\sigma_y}{\sqrt{2}}, \frac{\sigma_z}{\sqrt{2}} \bigg\}.
\end{equation}
Density matrices are then written as $d^2$ vectors identified by double bra or ket brackets
\begin{equation}
\label{eq:rho_HS}
    \vert\rho\rangle\rangle = \sum_k \vert P_k \rangle \rangle \langle\langle P_k \vert \rho \rangle \rangle = \sum_k \vert P_k \rangle \rangle \; \text{tr}\lbrace P_k^\dagger \rho\rbrace,
\end{equation}
whereas the Hilbert-Schmidt inner product is given by
\begin{equation}
\label{eq:HS_inner_prod}
    \langle\langle \alpha \vert \beta \rangle \rangle = \text{tr}\lbrace \alpha^\dagger \beta\rbrace \; .
\end{equation}
Operators describing linear maps can be written in the HS space as $d^2 \times d^2$ operators
\begin{equation}
\label{eq:op_HS}
    \mathcal{O}_\Lambda = \sum_{jk} \vert P_j \rangle\rangle \langle \langle P_j \vert \hat{\mathcal{O}}_\Lambda \vert P_k \rangle\rangle \langle\langle P_k \vert
\end{equation}
with $\langle \langle P_j \vert \hat{\mathcal{O}}_\Lambda \vert P_k \rangle\rangle = \text{tr}\lbrace P_j \Lambda\left(P_k\right) \rbrace$.
In our case, the linear map is simply given by the corresponding unitary pulse gate, $\Lambda\left(P_k\right) = UP_k U^\dagger$. \\ 
A measurement of a POVM $\langle\langle E\vert$ of the action of the gate $G$ on the initial state $\vert\rho\rangle\rangle$ is then given by
\begin{equation}
\label{eq:state_evolution}
    p_G = \langle\langle E\vert G \vert \rho \rangle\rangle.
\end{equation}
%


\section{Fluence}
\label{sec:fluence}

\begin{figure}
    \centering
    \includegraphics[width=0.48\textwidth]{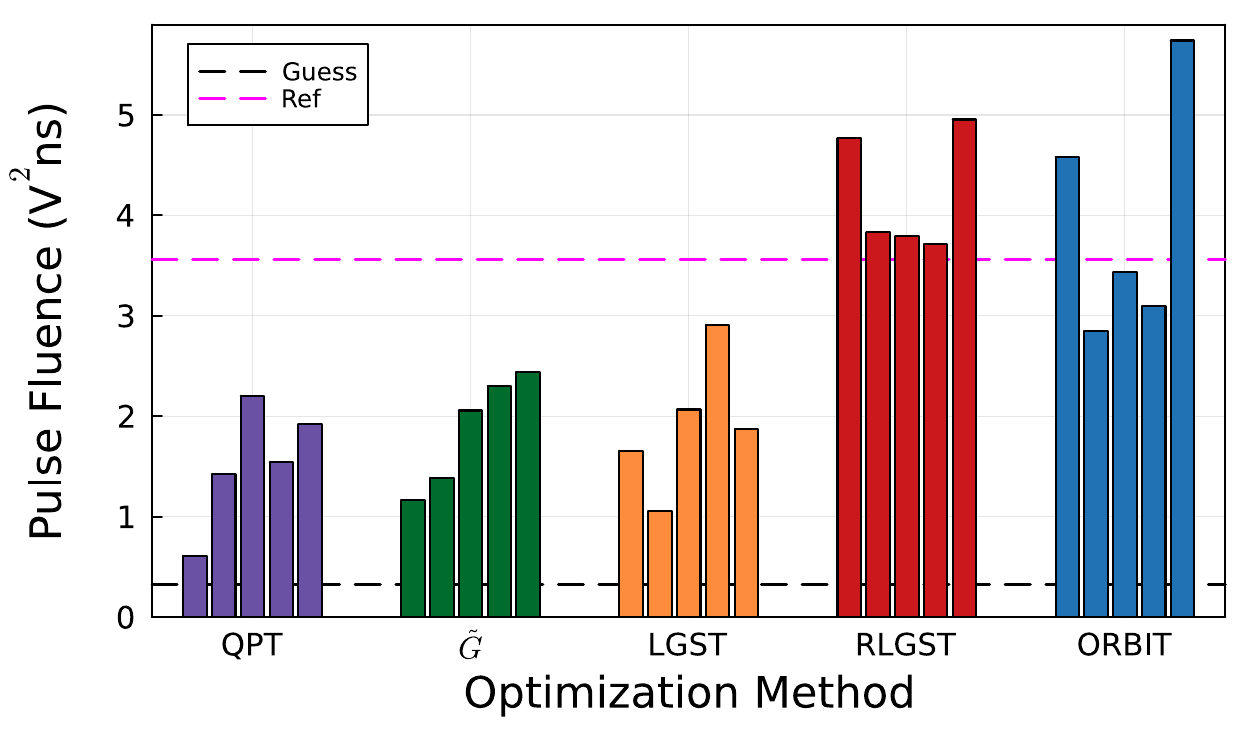}
    \caption{Fluence of the optimized pulses. We perform five optimization runs per method shown in the x-axis. The fluence of the final optimized pulse is shown by the height of the individual bars. The black dashed line corresponds to the fluence of the guess pulse and the magenta one to the fluence of the reference pulse.}
    \label{fig:fluence}
\end{figure}

To better understand why certain methods achieve significantly higher and consistent gains than others, we calculate the fluence of our optimized pulses.
The fluence is given by 
\begin{equation}
\label{eq:fluence}
    \Gamma=\int_0^{t_p} \left( a_x^2\left(t\right) + a_y^2\left(t\right)\right) \text{d}t,
\end{equation}
with the time-dependent amplitudes $a_x^2\left(t\right)$ and $a_y^2\left(t\right)$ of Eq.~(3) from the main text and the pulse length $t_p$.
While we ensure that at no point any pulse can have an amplitude larger then the reference pulse, the optimized pulse can achieve a higher fluence due to its longer pulse length. \\
The results for each optimization are shown in Fig.~\ref{fig:fluence}, where the height of the bars denote the fluence of the corresponding optimized pulse.
For RLGST and ORBIT we observe on average a slightly higher fluence than for the reference pulse, while the fluence of all other methods is located between the one of the reference and the guess pulse.
To achieve their enhanced robustness for any investigated method, RLGST and ORBIT thus require a significantly higher fluence than the other methods.
The increased fluence seems to be a perquisite for the gate-set's successful application at long time-scales, which is in line with the observations in the main text. \\
The optimized pulses of $\Tilde{G}$ show a much smaller fluence compared to the reference pulse.
Yet, the method shows a significantly higher gain than the reference pulse when evaluated by itself.
This makes the method particularly suitable if heating imposes an experimental limitation and the gate-set is to be optimized for specific applications of the gates to arbitrary basis states.

\section{Maximum-Likelihood Estimation}

\begin{figure}
    \centering
    \includegraphics[width=0.49\textwidth]{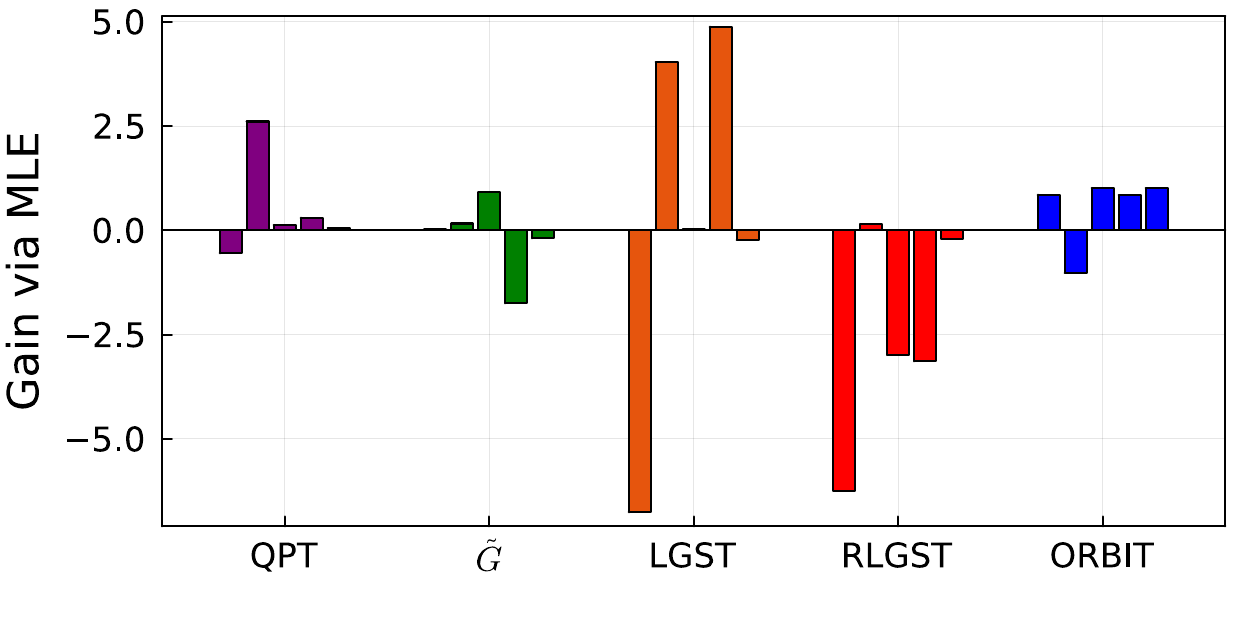}
    \caption{Evaluation of the different optimization runs through MLE. The abscissa shows the chosen optimization method, the ordinate the achieved gain.}
    \label{fig:mle_gain}
\end{figure}

Typically, the estimates provided by LGST are used as starting points for a maximum-likelihood estimation (MLE) to obtain real, physical estimates of the applied gates and states.
For such an MLE we take the expectation values $p_{ijk}$ (see Eq.~(7) in the main text) measured with LGST and minimize
%
\begin{equation}
    \sum_{ijk} \biggl( p_{ijk} - \sum_{mnrstu} \left( \chi_{F_i} \right)_{tu} \left( \chi_{G_k} \right)_{rs} \left( \chi_{F_j} \right)_{mn} \text{tr} \{ E P_t P_r P_m \rho P_n P_s P_u \}  \biggl)
\end{equation}
constrained by
\begin{equation}
\begin{split}
    &\sum_{mn} \left( \chi_{G} \right)_{mn} \text{tr} \left( P_m P_r P_n \right) - \delta_{0r} = 0 \, , \; \forall \; G \in \mathcal{G} \\
    &\text{tr} \{ \rho \} = 1 \, , \\
    &\mathds{1} - E \succcurlyeq 0
\end{split}
\end{equation}
according to Ref.~\cite{greenbaum2015introduction}. 
The process matrix $\chi_{G}$ describes the action of a gate on a given density matrix by
\begin{equation}
    \Lambda(\rho) = \sum_{i, j = 1}^{d^2} \left( \chi_{G} \right)_{ij} \, P_i \rho P_j
\end{equation}
with the Pauli operators $P_i$, $P_j$.
Performing the MLE for $20$ repeated measurements of a fixed set of pulses results in a variance of about $ 1 \%$ for the obtained process matrix fidelities.
From the LGST estimates of the gates one can calculate estimates for the process matrices, which are then used as a starting point for the minimization.
We perform the MLE for the five optimization runs per method. 
The process matrices resulting from the MLE are compared to the target process matrices to determine their fidelity.
Using the sum of the individual fidelities to calculate the optimization gain according to Eq.~(2) in the main text leads to the results shown in Fig.~\ref{fig:mle_gain}. 
We observe a huge variance between different optimization runs for one method, as well as extremely large and small gain values.
If we use the target process matrices as a starting point for the minimization such that the LGST estimates do not negatively influence the minimization, we obtain a similar result.
This is in stark contrast to all other evaluation methods and is reminiscent of the results from the evaluation with LGST.
The measurement errors of our experiment seem to strongly influence the MLE and do not allow us to to clearly distinguish between the gate-set of the guess, reference and optimized pulse.
For this reason, the MLE is omitted as evaluation method.
%

\section{Randomized benchmarking experiments}

\begin{figure}
    \centering
    \includegraphics[width=0.49\textwidth]{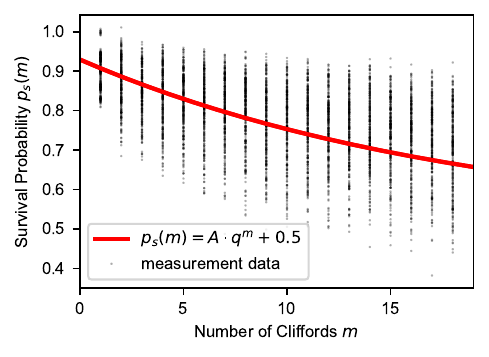}
    \caption{Randomized benchmarking experiment for an optimized gate-set. The abscissa shows the number of applied Clifford gates $m$ and the ordinate shows the corresponding survival probability $p_{s}\left(m\right)$. The data is fitted using a single exponential decay illustrated by the red line.}
    \label{fig:rb_example}
\end{figure}
%


We additionally use randomized benchmarking (RB)~\cite{knill2008randomized, magesan2011scalable, magesan2012characterizing} to evaluate our optimized gate-sets performance.
An exemplary measurement is shown in Fig.~\ref{fig:rb_example}, where the gate-set was optimized using ORBIT.
Heating limits us to a maximum circuit length of $18$ Cliffords and we average over $300$ randomized circuits for each circuit length.
The survival probability $p_{s}\left(m\right)$ is fitted with
\begin{equation}
    p_{s}\left(m\right) = A \cdot q^m+B,
    \label{eq:fit_function}
\end{equation}
where $m$ corresponds to the number of Cliffords.
Due to our normalization, all SPAM errors are absorbed by $A$, leading to $B=0.5$ by definition.
%
%
The parameter $q$ is used to calculate the average error per Clifford according to Ref.~\cite{magesan2011scalable, magesan2012characterizing}.
For the shown example we obtain an average error rate per Clifford of $r=0.0258\pm0.0007$ with $A=0.430\pm0.008$.

\newpage
\section{Optimization Metrics}
\label{sec:optimization_metrics}

Tab.~\ref{tab:optimization_metrics} shows the average number of evaluation steps of our optimizations for the selected method.
RLGST and ORBIT require notably more evaluation steps than an optimization with QPT, $\Tilde{G}$ or LGST.
The parameter landscape of those two methods must thus be significantly more complex, requiring more evaluation steps until the optimizer converges according to the set stopping criteria. \\
In addition, the average duration of one evaluation step for ORBIT and RLGST is also significantly longer than for the other methods.
To obtain the corresponding FoM of the two methods, we average over 300 circuits, i.e. we perform a measurement sequence which contains 236 more measurements than e.g. for $\Tilde{G}$.
These additional measurements increase the overall length of the measurement sequence, leading to a longer uploading time to the AWG and thus to a longer mean evaluation step duration.

\begin{table}[htb]
    \centering
    \caption{Optimization Metrics. The average number of evaluation steps, their average duration and the average total optimization time per method is displayed together with the corresponding uncertainty.}
    \begin{tabular}{|c|c|c|c|}
         \hline
         method & mean number of evaluation steps & mean evaluation step duration (s) & mean optimization duration (h)\\
         \hline\hline
         QPT            & $872 \pm 26$          & $40.524 \pm 0.028$        & $10.3 \pm 0.4$ \\
         $\Tilde{G}$    & $880 \pm 80$          & $46.74 \pm 0.17$          & $14.1 \pm1.8$ \\
         LGST           & $723 \pm 9$           & $47.25 \pm 0.16$          & $11.5 \pm0.8$ \\
         RLGST          & $1220 \pm 60$         & $66.49 \pm 0.25$          & $32 \pm5$ \\
         ORBIT          & $1100 \pm 100$        & $67.3 \pm 0.4$            & $28.0 \pm2.9$ \\
         \hline
    \end{tabular}
    \label{tab:optimization_metrics}
\end{table}

\section{FoM progression during the optimizations}

\begin{figure}
    \centering
    \begin{tikzpicture}
    \node (image) at (-9,0) {\includegraphics[width=0.49\textwidth]{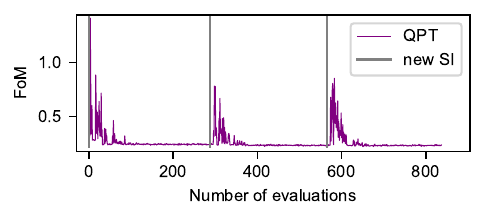}};
    \node (image) at (0,0) {\includegraphics[width=0.49\textwidth]{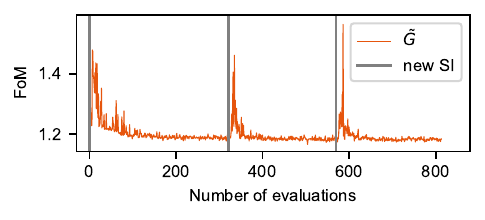}};
    \node (image) at (-9,-4) {\includegraphics[width=0.49\textwidth]{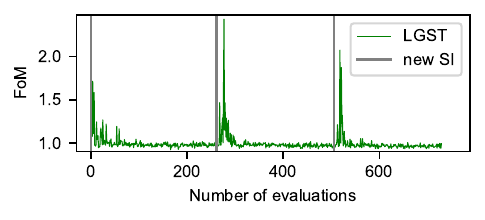}};
    \node (image) at (0,-4) {\includegraphics[width=0.49\textwidth]{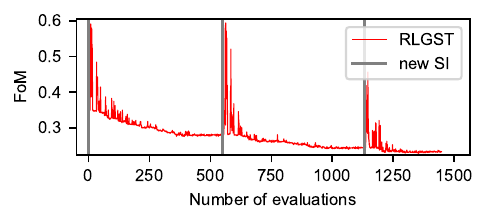}};
    \node (image) at (-9,-8) {\includegraphics[width=0.49\textwidth]{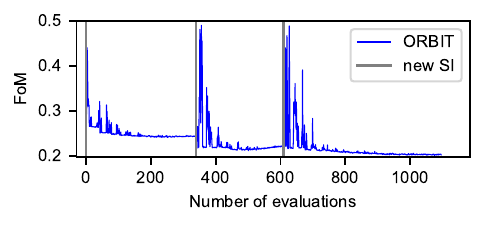}};

    \node [anchor=east] at (-12.65, 2) {\large \textbf{a)}};
    \node [anchor=east] at (-3.65, 2) {\large \textbf{b)}};
    \node [anchor=east] at (-12.65, -2) {\large \textbf{c)}};
    \node [anchor=east] at (-3.65, -2) {\large \textbf{d)}};
    \node [anchor=east] at (-12.65, -6) {\large \textbf{e)}};
    \end{tikzpicture}
    \caption{FoM progression.
    Exemplary FoM progression during an optimization via \textbf{a)} QPT, \textbf{b)} $\Tilde{G}$, \textbf{c)} LGST, \textbf{d)} RLGST and \textbf{e)} ORBIT.
    The abscissa shows the number of evaluations and the ordinate the absolute value of the corresponding FoM.
    The grey lines mark the beginning of a new super-iteration.}
    \label{fig:fom_progression}
\end{figure}

Fig.~\ref{fig:fom_progression} shows the FoM progression during an optimization for the different analysis methods.
The optimization via QPT in Fig.~\ref{fig:fom_progression}~a) shows a clear minimization of the FoM.
We observe a strong modulation of the FoM during the beginning of a new super-iteration.
In addition, the FoM converges very quickly and shows almost no variance between different super-iterations. \\
For the optimization via $\Tilde{G}$ in Fig.~\ref{fig:fom_progression}~b) we also see a clear minimization but the ratio between the final FoM and the one of the initial guess is small compared to the other methods.
Depending on the optimization run we can observe a further decrease of the FoM through additional super-iterations. \\
For LGST in Fig.~\ref{fig:fom_progression}~c) we observe again strong variations of the FoM at the beginning of a super-iteration.
The improvement of the FoM is almost within the noise, due to the problems discussed in the main text.
QuOCS is well equipped for such a task through the use of re-evaluation steps to correctly determine if the FoM truly improved or not.
This allows us to enhance the gate-set's performance even for LGST. \\
For RLGST in Fig.~\ref{fig:fom_progression}~d) and ORBIT in Fig.~\ref{fig:fom_progression}~e) we observe a very clear minimization of the FoM from the initial guess. 
The FoM seems to be well defined such that QuOCS can easily improve the pulse shape and converge within the set boundaries. 
One can see a clear improvement of the FoM through the use of additional super-iterations.
As we limit ourselves to three super-iterations due to time-limitations, we cannot exclude that the gains reported in the main text could be much higher for longer optimizations.


\bibliography{references}